\documentstyle[preprint,aps,floats]{revtex}
\input epsf.tex
\def\DESepsf(#1 width #2){\epsfxsize=#2 \epsfbox{#1}}
\begin{document}
\preprint{\vbox{\hbox{}}}
\draft

\title{
Constraints on SUSY Gluonic Dipole Interaction\\
from $B\to K \pi$ Decays
}
\vfill
\author{Xiao-Gang He, Jenq-Yuan Leou and Jian-Qing Shi}
\address{
\rm Department of Physics, National Taiwan University,
Taipei, Taiwan 10764, R.O.C.}

%
%
\vfill
\maketitle
\begin{abstract}
In low energy SUSY theories exchange of gluino and squark with
left-right mixing can produce a large gluonic dipole
interaction. In this paper we study the effects of this interaction
on $B\to K \pi$ using QCD improved factorization
method. The Standard Model predicts a smaller branching ratio for
$B^0 \to \bar K^0 \pi^0$ than experimental measured one.
We find that within the parameter space allowed from $B\to \gamma X_s$
constraint, the SUSY dipole
interaction can enhance this branching ratio to agree with the experimental
measurement. Combining recent data for all the four $\bar B^0
\to K^- \pi^+, \bar K^0 \pi^0$ and $B^- \to K^- \pi^0, \bar K^0 \pi^-$
decay modes, we find that
the allowed parameter space is reduced significantly
compared with that using $B\to X_s \gamma$ data alone.
It is found that
even with these constraints, the predictions for CP violation in these modes
can be dramatically different from those of the SM predictions.

\end{abstract}

\newpage

\section{Introduction}

There have been considerable experimental and theoretical efforts to
understand the properties of B systems. These studies have provided important
information about B decays and CP violation.
At the quark level the relevant Hamiltonian for B decays in the Standard Model
(SM) is well understood.
When going beyond the SM, there are new contributions. These new contributions
can modify or even completely change the SM predictions\cite{1,2,2a}.
In SUSY theories with flavor changing interaction in the squark sector, it
is possible to generate large effects on hadronic $B$ decays while
their effects on other processes are small\cite{2,2a}.
In particular exchanges of gluino
and squark with left-right mixing can enhance the gluonic dipole
interactions of the forms
$\bar q\sigma^{\mu\nu}G^{\mu\nu}(1\pm \gamma_5)b$
by a large ratio factor of gluino mass $m_{\tilde g}$ to b quark mass $m_b$,
$m_{\tilde g}/m_b$,
compared with the SM prediction. Due to this
enhancement factor, even a tiny coefficient for the associated flavor changing
squark-gluino-quark interaction, a large gluonic dipole interaction can be
generated.

A large gluonic dipole interaction
can affect $B$ decays significantly.
It may help to explain the large branching ratios observed for $B\to X_s \eta'$,
although theoretical understanding is poor\cite{3,4a}.
It can also change theoretical predictions for other charmless hadronic
$B$ decays, such as $B\to K \pi, \pi\pi, \phi K$\cite{4a,4}. Using the recently
measured branching ratios for $B\to K\pi, \pi\pi$, one may be able to
constraint the allowed parameter space which can generate a large gluonic dipole
interactions and to provide interesting information about models beyond the
SM. In SUSY models
the same left-right squark mixing parameters
can also generate a photonic dipole interaction which can induce
$B\to X_s \gamma$ and $B\to X_d \gamma$. At present
$B\to X_s \gamma$ has been observed, but not $B\to X_d \gamma$.

In this paper we study in detail constraints on SUSY gluonic dipole
interaction using data from $\bar B^0 \to K^-\pi^+,
\bar K^0 \pi^0$ and $B^- \to K^- \pi^0, \bar K^0 \pi^-$ and also the
measured branching ratio of $B\to X_s \gamma$.
Previous studies of gluonic dipole interactions on $B$ decays were based on
naive factorization approximation. Here we will use QCD improved factorization
method\cite{8,9}
which improves the analysis on several aspects, such as the number of
color, the gluon virtuality, the renormalization scale, scheme dependencies
and etc.
We find that the recent data on $B\to K \pi$ decays
can reduce, significantly, the allowed parameter space compared with using
$B\to X_s \gamma$ data alone,
while still allow large deviations from the SM predictions for
branching ratios and
CP asymmetries in these decay modes.

The paper is arranged as follows. In section II, we discuss gluonic
dipole interactions in low energy SUSY model with left-right squark mixing,
and obtain the decay amplitudes for $B\to K \pi$ decays using QCD improved
factorization. In section III, we first update constraints on the SUSY
gluonic dipole interactions from $B\to X_s \gamma$, and then study constraints
from $B\to K\pi$ decays. In section IV, we study CP rate asymmetry in
$B\to K \pi$. And in section V, we draw our conclusions.

\section{ SUSY gluonic dipole contributions to $B\to K \pi$}

In the SM the effective Hamiltonian for charmless photonic and
hadronic B decays with
$\Delta S = 1$ at the quark level is given by

\begin{eqnarray}
H_{eff} &=&{G_F\over \sqrt{2}}
\left \{ V_{ub} V^*_{us}(c_1 O_1 +c_2 O_2+ \sum^{12}_{i=3}c_iO_i)+
V_{cb} V^*_{cs} \sum_{i=3}^{12} c_i O_i\right \}.
\label{hamiltonian}
\end{eqnarray}
Here $O_i$ are quark, gluon and photon operators and are given by

\begin{eqnarray}
&&O_1 = (\bar s_i u_i)_{V-A} (\bar u_j b_j)_{V-A},\;\;
O_2 = (\bar s_i u_j)_{V-A} (\bar u_i b_j)_{V-A},\nonumber\\
&&O_{3(5)} = (\bar s_i b_i)_{V-A}\sum_{q'}
(\bar q^\prime_j q^\prime_j)_{V-(+)A},\;\;
O_{4(6)} = (\bar s_i b_j)_{V-A}\sum_{q'}
(\bar q^\prime_j q^\prime_i)_{V-(+)A},\nonumber\\
&&O_{7(9)} = {3\over 2}(\bar s_i b_i)_{V-A}\sum_{q'}
e_{q^\prime}(\bar q^\prime_j q^\prime_j)_{V+(-)A},\;\;
O_{8(10)} ={3\over 2} (\bar s_i b_j)_{V-A}\sum_{q'}
e_{q^\prime}(\bar q^\prime_j q^\prime_i)_{V+(-)A},\nonumber\\
&&O_{11} = {g_s\over 8\pi^2} m_b \bar s_i \sigma^{\mu\nu} G_{\mu\nu}^a
T^{ij}_a(1+\gamma_5)b_j,\;\;
O_{12} = {e\over 8\pi^2} m_b \bar s_i \sigma^{\mu\nu} F_{\mu\nu}
(1+\gamma_5)b_i,
\end{eqnarray}
where $(V\pm A)(V\pm A) =\gamma^\mu(1\pm\gamma_5) \gamma_\mu(1\pm \gamma_5)$,
$q^\prime = u,d,s,c,b$, $e_{q^\prime}$ is the electric charge number
of the $q^\prime$ quark, $T_a$ is the color SU(3) generator normalized as
$Tr(T^a T^b) = \delta^{ab}/2$,
$i$ and $j$ are color indices, and
$G_{\mu\nu}$ and $F_{\mu\nu}$ are the gluon and photon field strengths.

The Wilson coefficients $c_i$ have been calculated in
different schemes\cite{6}. In this paper we will use consistently the NDR
scheme. The values of $c_i$ at $\mu \approx m_b$ with
QCD corrections at NLO are given by \cite{8}

\begin{eqnarray}
&&c_1 = 1.081,\;\;c_2 = -0.190,\;\;c_3=0.014,\;\;c_4 = -0.036,\;\;
c_5=0.009,\;\;c_6 = -0.042,\nonumber\\
&&c_7= -0.011\alpha_{em},\;\;c_8=0.060\alpha_{em},\;\;
c_9=-1.254\alpha_{em},\;\;c_{10}=0.223\alpha_{em}.
\end{eqnarray}
And at LO are given by \cite{8}
\begin{eqnarray}
&&c_1 = 1.117,\;\;c_2 = -0.268,\;\;c_3=0.012,\;\;c_4 = -0.027,\;\;
c_5=0.008,\;\;c_6 = -0.034,\nonumber\\
&&c_7= -0.001\alpha_{em},\;\;c_8=0.029\alpha_{em},\;\;
c_9=-1.276\alpha_{em},\;\;c_{10}=0.288\alpha_{em},\nonumber\\
&&c_{11}=-0.151,\;\;c_{12}=-0.318.\label{wilsonC}
\end{eqnarray}
Here $\alpha_{em}=1/129$ is the electromagnetic fine structure constant.
We use LO Wilson coefficients with the matrix elements enter the decay amplitudes
at next-to-leading order.

In SUSY models,
exchanges of gluino and squark with left-right squark mixing, can generate
a large contribution to $c_{11,12}$ at one loop level.
In general these contributions can generate
a gluonic dipole interaction with the same chirality as the SM one, as well as
with an opposite chirality as the SM one, that
is, an interaction similar to $O_{11,12}$ but with $1+\gamma_5$
replaced by $1-\gamma_5$. It is difficult to
carry out an analysis in the full parameter space.
We will first consider the new contributions with the same chirality
as the SM one setting the opposite chirality one to zero,
and then the opposite case.

The effective Wilson coefficient for $c^{susy}_{11,12}$ obtained in the mass insertion
approximation is given by, for the case with the
same chirality as the SM ones\cite{7},

\begin{eqnarray}
&&c_{11}^{susy}(m_{\tilde g}) = {\sqrt{2}\pi \alpha_s(m_{\tilde g})\over G_F
m^2_{\tilde g}}
{\delta_{LR}^{bs}\over V_{tb}V_{ts}^*}{m_{\tilde g}\over m_b}
G_0(x_{gq}),\nonumber\\
&&c_{12}^{susy}(m_{\tilde g}) = {\sqrt{2}\pi \alpha_s(m_{\tilde g})\over
G_Fm^2_{\tilde g}}
{\delta_{LR}^{bs}\over V_{tb}V_{ts}^*}{m_{\tilde g}\over m_b}
F_0(x_{gq}),\nonumber\\
&&G_0(x) = {x\over 3(1-x)^4}[22-20x-2x^2+16x \ln(x)-x^2 \ln(x)+9\ln(x)],\nonumber\\
&&F_0(x) = -{4x\over 9(1-x)^4} [1+4x-5x^2+4x\ln(x) +2x^2\ln(x)], \label{csusymg}
\end{eqnarray}
where $\delta_{LR}^{bs}$ is the mixing parameter of left and right
squarks, $x_{gq} = m^2_{\tilde g}/m^2_{\tilde q}$ is the ratio of
gluino and squark mass.

The coefficients $c^{susy}_{11,12}(m_{\tilde g})$ at $m_b$ are given by\cite{7}

\begin{eqnarray}
c^{susy}_{11}(\mu) = \eta c_{11}^{susy}(m_{\tilde g}),\;\;
c^{susy}_{12}(\mu) = \eta^2 c^{susy}_{12}(m_{\tilde g})+
{8\over 3}(\eta -\eta^2)c^{susy}_{11}(m_{\tilde g}), \label{csusymb}
\end{eqnarray}
where $\eta = (\alpha_s(m_{\tilde g})/\alpha_s(m_t))^{2/21}
(\alpha_s(m_t)/\alpha_s(mb))^{2/23}$.

In the SM, $c_{11,12}$ are proportional to $m_b/m_W^2$. From the above
expressions, it is clear that the SUSY contributions are proportional to
$1/m_{\tilde g}$. If $m_{\tilde g}$ is of order a few hundred GeV, there is
an enhancement factor of $m_{\tilde g}/m_b (m_W^2/m_{\tilde g}^2)$ for the
SUSY gluonic dipole
interaction. In this case even a small $\delta_{LR}^{bs}$ can have a large
effect on $B$ decays.

To obtain $c^{susy}_{11,12}$ for opposite chirality case, one
just adds in two more operators similar to $O_{11,12}$ but with
$1+\gamma_5$ replaced by $1-\gamma_5$ and the Left-Right mixing
parameter $\delta_{LR}^{bs}$ replaced by the Right-Left mixing
parameter $\delta^{bs}_{RL}$.

We follow Ref.\cite{8} to obtain the $B\to K\pi$ decay amplitudes. They are
given by

\begin{eqnarray}
&&A(\bar B^0\to \bar K^0 \pi^0)
={G_F\over 2} if_{\pi} (m^2_B-m^2_K)F_0^{B\to K}(m^2_\pi)
[V_{ub}V_{us}^*(a_2 + {3\over 2}(-a_7 + a_9))
+V_{cb}V_{cs}^*{3\over 2}(-a_7+a_9)]\nonumber\\
&&+{G_F\over 2} i f_K (m^2_B-m^2_\pi)F_0^{B\to \pi}(m^2_K)
[V_{ub}V_{us}^*(-a_4^u +{1\over 2} a_{10}^u - R_K (a_6^u - {1\over 2} a_8^u)
)\nonumber\\&&
+V_{cb}V_{cs}^*(-a^c_4+{1\over 2} a^c_{10}-R_K(a^c_6 - {1\over 2}a_8^c)
]-{G_F\over 2} if_B f_\pi f_K (V_{ub}V_{us}^*+V_{cb}V_{cs}^*)(b_3 - {1\over 2}b_3^{EW}),\nonumber\\
\nonumber\\
&&A(\bar B^0\to K^- \pi^+)
={G_F\over 2} if_{\pi} (m^2_B-m^2_K)F_0^{B\to K}(m^2_\pi)
[V_{ub}V_{us}^*(a_1+a_4^u +a_{10}^u+ R_K(a^u_6+a_8^u))\nonumber\\
&&+V_{cb}V_{cs}^*(a_4^c+a_{10}^c + R_K(a^c_6+a_8^c)]+{G_F\over \sqrt{2}} if_B f_\pi f_K (V_{ub}V_{us}^*+V_{cb}V_{cs}^*)(b_3 - {1\over 2}b_3^{EW}),\nonumber\\
\nonumber\\
&&A(B^-\to K^- \pi^0)
={G_F\over 2} if_{K} (m^2_B-m^2_\pi)F_0^{B\to \pi}(m^2_\pi)
[V_{ub}V_{us}^*(a_2 +{3\over 2}(-a_7 + a_9))
+V_{cb}V_{cs}^*{3\over 2}(-a_7+a_9))]\nonumber\\
&& +
{G_F\over 2} i f_K (m^2_B-m^2_\pi)F_0^{B\to \pi}(m^2_K)
[V_{ub}V_{us}^*(a_2+a_4^u + a_{10}^u+ R_K (a_6^u + a_8^u))\nonumber\\
&&+V_{cb}V_{cs}^*(a^c_4+R_K(a^c_6 +a_8^c +a_{8a}^c)
+ a^c_{10})]+{G_F\over \sqrt{2}} if_B f_\pi f_K [V_{ub}V_{us}^*(b_2+b_3+b_3^{EW})+V_{cb}V_{cs}^*(b_3 +b_3^{EW})],\nonumber\\
\nonumber\\
&&A(B^-\to \bar K^0 \pi^-) =
{G_F\over 2} i f_K (m^2_B-m^2_\pi)F_o^{B\to \pi}(m^2_K)
[V_{ub}V_{us}^*(a_4^u -{1\over 2} a_{10}^u+ R_K (a_6^u -{1\over 2} a_8^u ))\nonumber\\
&&+V_{cb}V_{cs}^*(a^c_4+R_K(a^c_6 -{1\over 2} a_8^c )
-{1\over 2} a^c{10})]+{G_F\over \sqrt{2}} if_B f_\pi f_K [V_{ub}V_{us}^*(b_2+b_3+b_3^{EW})+V_{cb}V_{cs}^*(b_3 +b_3^{EW})].
\end{eqnarray}
Here $R_K = 2m^2_K/m_s m_b$. $a_i$ and $b_i$ coefficients are related to the
Wilson coefficients. In the above we have neglected small contributions from
$O_{12}$.
Including the lowest $\alpha_s$ order corrections, $a^q_i$
are given by

\begin{eqnarray}
&&a_1 = c_1 +{c_2\over N}\left[1 + {C_F \alpha_s \over 4\pi} V_K \right]+\frac{c_2}{N_c}\frac{C_F \pi \alpha_s}{N_c}H_{K\pi},\nonumber\\
&&a_2 = c_2 +{c_1\over N}\left[1 + {C_F \alpha_s \over 4\pi} V_\pi \right]+\frac{c_1}{N_c}\frac{C_F \pi \alpha_s}{N_c}H_{\pi K},\nonumber\\
&&a_4^p = c_4 +{c_3\over N}\left[1 + {C_F \alpha_s \over 4\pi} V_K \right]+ {C_F \alpha_s \over 4\pi} {P^p_{K,2}\over N_c}+\frac{c_3}{N_c}\frac{C_F \pi \alpha_s}{N_c}H_{K\pi },\nonumber\\
&&a_6^p = c_6 +{c_5\over N}\left[1 -6\cdot{C_F \alpha_s \over 4\pi} \right]+ {C_F \alpha_s \over 4\pi} {P^p_{K,3}\over N_c},\nonumber\\
&&a_7 = c_7 +{c_8\over N}\left[1 + {C_F \alpha_s \over 4\pi} (-V'_\pi)\right]+\frac{c_8}{N_c}\frac{C_F \pi \alpha_s}{N_c}(-H'_{\pi K}),\nonumber\\
&&a_8^p = c_8 +{c_7\over N}\left[1 -6\cdot {C_F \alpha_s \over 4\pi} \right]+ {\alpha_{em} \over 9\pi} {P^{p,EW}_{K,3}\over N_c},\nonumber\\
&&a_9 = c_9 +{c_{10}\over N}\left[1 + {C_F \alpha_s \over 4\pi} V_\pi\right]+\frac{c_{10}}{N_c}\frac{C_F \pi \alpha_s}{N_c}H_{\pi K},\nonumber\\
&&a_{10}^p = c_{10} +{c_9\over N}\left[1 + {C_F \alpha_s \over 4\pi} V_K \right]+ {\alpha_{em} \over 9\pi} {P^{p,EW}_{K,2}\over N_c}+\frac{c_9}{N_c}\frac{C_F \pi \alpha_s}{N_c}H_{K\pi },
\end{eqnarray}
where $C_F = (N_c^2-1)/(2N_c)$, and $N_c=3$. The quantities $V_M^{(\prime)}$,
$H^{(\prime)}_{M_2 M_1}$, $P^p_{K,2}$, $P^p_{K,3}$, $P^{p,EW}_{K,2}$ and $P^{p,EW}_{K,3}$
are hadronic parameters that contain all nonperturbative dynamics.

The vertex corrections $V_M^{(\prime)}$(M=$\pi$,K) are give by
\begin{eqnarray}
V_M &=& 12 \ln{m_b \over \mu} - 18 + \int^1_0 dx g(x) \phi_M(x),\nonumber\\
V'_M &=& 12 \ln{m_b \over \mu} - 18 + \int^1_0 dx g(1-x) \phi_M(x),\nonumber\\
g(x)&=&  3{1-2x\over 1-x} \ln x -3i\pi\nonumber\\
&&+\left[ 2 \mbox{Li}_2(x)-\ln^2 x +\frac{2\ln x}{1-x}-(3+2i \pi)-(x\leftrightarrow1-x)\right],
\end{eqnarray}
where $\mbox{Li}_2(x)$ is the dilogarithm. $\phi_M(x)$ is the leading-twist
light cone distribution amplitude.
This distribution amplitude can be expansion in Gegenbauer ploynomials. We truncate this
expansion at $n=2$.
\begin{eqnarray}
\phi_M(x,\mu) = 6x(1-x)\left[ 1+\alpha_1^M(\mu)C_1^{(3/2)}(2x-1)
+\alpha_2^M(\mu) C_2^{(3/2)(2x-1)}\right],
\end{eqnarray}
where $C_1^{(3/2)}(u)=3u$ and $C_2^{(3/2)}(u)= \frac{3}{2}(5u^2-1)$.
The distribution amplitude parameters $\alpha_{1,2}^M$ for $M=K,\pi$ are:
$\alpha_1^K=0.3$, $\alpha_2^K=0.1$, $\alpha_1^\pi=1$ and $\alpha_2^\pi=0.1$.

The penguin contributions $P^p_{K,2},\;P^{p}_{K,3},\;P^{p,EW}_{K,2}$ and $P^{p,EW}_{K,3}$ are given by
\begin{eqnarray}
P^p_{K,2}&=&c_1 \left[ \frac{4}{3}\ln \frac{m_b}{\mu}+\frac{2}{3}-G_K(s_p)\right]+
c_3 \left[ \frac{8}{3}\ln \frac{m_b}{\mu}+\frac{4}{3}-G_K(0)-G_K(1)\right] \nonumber \\
&&+(c_4+c_6) \left[ \frac{4 n_f}{3}\ln \frac{m_b}{\mu}-(n_f-2)G_K(0)-G_K(s_c)-G_K(1)\right]\nonumber \\
&&-2c_{11} \int^1_0 \frac{dx}{1-x}\phi_K(x),\nonumber \\
P^{p,EW}_{K,2}&=&(c_1+N_c c_2) \left[ \frac{4}{3}\ln \frac{m_b}{\mu}+\frac{3}{2}
-G_K(s_p)\right]-2c_{12} \int^1_0 \frac{dx}{1-x}\phi_K(x),\nonumber \\
P^{p}_{K,3}&=&c_1 \left[ \frac{4}{3}\ln \frac{m_b}{\mu}+\frac{2}{3}-\hat G_K(s_p)\right]+
c_3 \left[ \frac{8}{3}\ln \frac{m_b}{\mu}+\frac{4}{3}-\hat G_K(0)-\hat G_K(1)\right] \nonumber \\
&&+(c_4+c_6) \left[ \frac{4 n_f}{3}\ln \frac{m_b}{\mu}-(n_f-2)\hat G_K(0)-\hat G_K(s_c)-\hat G_K(1)\right]-2 c_{11},\nonumber \\
P^{p,EW}_{K,3}&=&(c_1+N_c c_2) \left[ \frac{4}{3}\ln \frac{m_b}{\mu}+\frac{3}{2}
-\hat G_K(s_p)\right]-3c_{12},
\end{eqnarray}
with
\begin{eqnarray}
G_K(s)&=&\int_0^1 dx G(s-i\epsilon,1-x)\phi_K(x), \nonumber\\
\hat G_K(s)&=&\int_0^1 dx G(s-i\epsilon,1-x)\phi^K_p(x), \nonumber\\
G(s,x)&=&-4\int_0^1 du u(1-u)\ln [s-u(1-u)x],
\end{eqnarray}
where $\phi_p^K(x)=1$.

The hard-scattering contributions $H_{\pi K},\;H'_{\pi K},$ and $H_{K\pi}$ contain many poorly know parameters,we follow \cite{8} use $H_{\pi K} =0.99$ and
\begin{eqnarray}
H'_{\pi K}=H_{\pi K},\;\;\;\; H_{K\pi}=R_{\pi K} H_{\pi K},
\end{eqnarray}
where
\begin{eqnarray}
R_{\pi K}\simeq\frac{F_0^{B\to K}(0)f_\pi}{F^{B\to \pi}_0(0)f_K}.
\end{eqnarray}
All scale-dependent quantities for hard spectator contributions are evaluated at $\mu_h=\sqrt{\Lambda_h m_b}$ with $\Lambda_h=0.5$GeV.

The annihilation coefficients are
\begin{eqnarray}
&&b_2=\frac{C_F}{N_c^2}c_2 A_1^i,\;\;\;\;b_3=\frac{C_F}{N_c^2}[ c_3 A^i_1+c_5 (A^i_3+A^f_3)+N_c c_6A_3^f],\nonumber \\
&&b_3^{EW}=\frac{C_F}{N_c^2} [c_9 A^i_1+c_7(A_3^i+A_3^f)+N_c c_8 A_3^f],\nonumber \\
&&b_4^{EW}=\frac{C_F}{N_c^2}[c_{10} A_1^i+c_8 A^i_2].
\end{eqnarray}
with
\begin{eqnarray}
&&A_1^i=A_2^i=\pi \alpha_s \left[ 18 \left(X_A-4+\frac{\pi^2}{3} \right)+2r_\chi^2 X_A^2 \right],\nonumber\\
&&A_3^f=12\pi \alpha_s r_\chi^2 (2X_A^2-X_A),
\end{eqnarray}
where $r_\chi=2 \mu_h /m_b$, $X_A=\int^1_0 dy/y$ parameterizes the divergent endpoint integrals, We use $X_A=\ln(m_B/ \Lambda_h)$.
All scale dependent quantities for annihilation contributions are evaluated at $\mu_h$.

In the numerical evaluation of these expressions we follow Ref. \cite{8} consistently drop high-order
terms in the products of the Wilson coefficients with the next-to-leading order
corrections. I.e., we evaluate all the corrections beyond naive factorization with leading
order Wilson coefficients.

For the case where the SUSY contributions have the same chirality as the
SM one, $c_{11} = c_{11}^{SM} + c_{11}^{susy}$.
For the opposite chirality case, one needs to change $c_{11}$ to
$c^{SM}_{11} - c_{11}^{susy}$
due to the replacement of $1+\gamma_5$ by $1-\gamma_5$.

\section{Constraints on $c_{11}$ from $B\to \gamma X_s$ and $B\to K \pi$}

\subsection{Constraints from $B\to X_s\gamma$}

\begin{figure}[htb]
\centerline{ \DESepsf(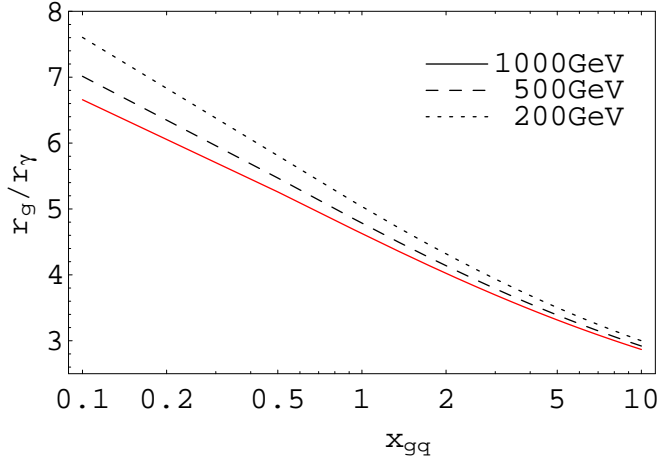 width 10cm)}
\smallskip
\caption {
Ratio of $r_g/ r_\gamma$ vs. $x_{gq}$.
The solid, dashed and dotted lines correspond to gluino mass of
$1000$ GeV, $500$ GeV and $200$ GeV, respectively.} \label{ratio}
\end{figure}

We now study the constraints on gluonic dipole interaction using $B\to X_s \gamma$.
To this end we define $r_\gamma =
|c_{12}^{susy} (m_b)/c_{12}^{SM}(m_b)
|$ and $r_g = |c_{11}^{susy}(m_b) /c_{11}^{SM}(m_b) |$.
$c_{11,12}^{SM}(m_b)$ denote SM contributions, which are given in the Eq.(\ref{wilsonC}).
Using Eq.(\ref{csusymb}) and replacing $c_{11,12}^{susy}$ with
Eq.(\ref{csusymg}), we obtain
\begin{eqnarray}
{r_g \over r_\gamma} = [\eta \frac{F_0(x_{gq})}{G_0(x_{gq})}+
\frac{8}{3}(1-\eta)]^{-1} \frac{c_{12}^{SM}}{c_{11}^{SM}}\;.\label{ratiogga}
\end{eqnarray}
$r_\gamma / r_g$ is not sensitive to gluino mass, but strongly
depends on $x_{gq}$.
In Fig.(\ref{ratio}) we show $r_g/r_\gamma$ as a function of $x_{gq}$ for different
gluino masses.
Note that both $r_g$ and $r_\gamma$ have a common CP violating phase $\delta$
which is equal to the phase difference of the SM phase
and the SUSY phase of $\delta_{LR}^{sb}$.

For a given $r_\gamma$, $r_g$ is known as a function of
$x_{gq}$. Therefore constraints on
$r_\gamma$ from $B\to X_s \gamma$ can be translated into constraint
on $r_g$. The constraints on $r_\gamma$ is given by\cite{2a,10}
\begin{eqnarray}
Br(B\to X_s \gamma) &=&Br(B \to X e \bar \nu _e) \frac{|V_{ts}^* V_{tb}|^2}{|V_{cb}|^2}
\frac{6 \alpha_{em}}{ \pi g(m_c/m_b) \eta} |c_{12}^{eff}(m_b)|^2 \;, \label{btosga}
\end{eqnarray}
where
$c_{12}^{eff}(M_b)=c_{12}^{SM}(m_b)(1+r_\gamma e^{i \delta})$,
$g(z)=1-8z^2+8z^6-z^8-24z^4 ln z$, and $\eta=1-2f(r,0,0) \alpha_s(m_b)/3\pi$
with $f(r,0,0)=2.41$\cite{10}.
Here we have used the leading order result. The use of next-leading order
result\cite{10a} will slightly change the results (at the level of $10\%$),
but will not change the main conclusions.

For the opposite chirality case, one needs to replace $|c_{12}^{eff}|^2
=|c^{SM}_{12}(1+r_\gamma e^{i \delta})|^2$ in the above
by $|c_{12}^{eff}|^2 = |c_{12}^{SM}|^2(1+r_\gamma^2)$
because the SM and SUSY contributions have different chiralities.
Note that in this case
$r_{\gamma, g}$ also have a common phase equal to the
phase difference of SM and SUSY due to $\delta_{RL}^{sb}$
which will be also indicated by $\delta$.

\begin{figure}[htb]
\centerline{ \DESepsf(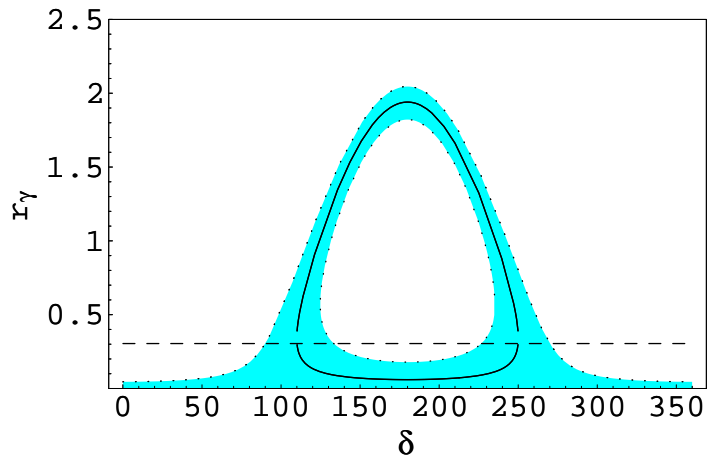 width 10cm)}
\smallskip
\caption {
Constraint on $r_\gamma$ from $B \to X_s \gamma$.
For the case with
the same chirality as the SM one,
the 95\% C.L. allowed region is indicated by the shaded region.
For the opposite chirality case,
the 95\% C.L. allowed region is below the dashed line.
} \label{rgamma}
\end{figure}

\begin{figure}[htb]
\centerline{ \DESepsf(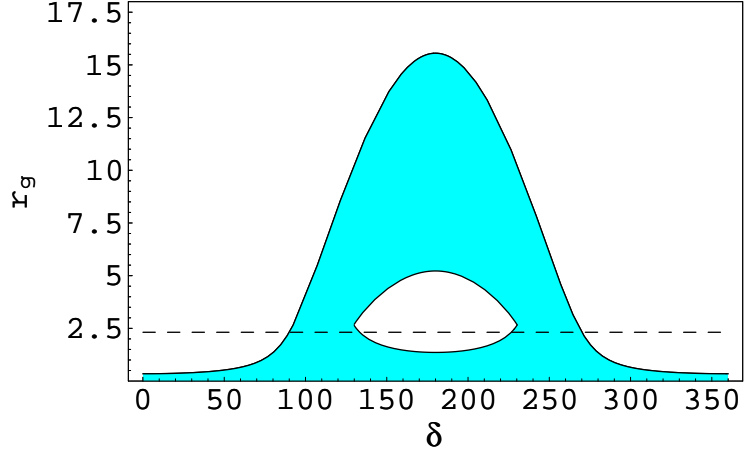 width 10cm)}
\smallskip
\caption {
Constraint of $r_g$ in $95\%$C.L., with $200 < m_{\widetilde{g}} <1000$, $0.1 < x_{gq} <10$.
The shaded region and the region below the dashed line are the allowed
regions with
the 95\% C.L. for the case with the same and opposite chiralities,
respectively. }\label{rgluon}
\end{figure}

In the numerical analysis we will use
$Br(B\to X e \bar \nu_e)=10.4\%$,
$m_b=4.2GeV$, $m_c=1.3 GeV$,
and $|V_{ts}^* V_{tb}|^2/ |V_{cb}|^2=0.95$.
Using the experimental allowed range $Br(B\to X_s \gamma) =(2.96 \pm
0.35)\times 10^{-4}$ averaged from CLEO, ALEPH, and Belle\cite{11},
one can obtain the allowed region for $r_\gamma$ by Eqs.(13) and
(14). The $95\%$ C.L. region of $r_\gamma$
is shown in Fig.(\ref{rgamma}). Combining information
from Figs.(\ref{ratio}) and (\ref{rgamma}), we obtain the allowed region for
$r_g$ in Fig.(\ref{rgluon}). We see that $c_{11}^{susy}$ can be
considerably larger than the SM. We note that the bounds of $r_\gamma$ and $r_g$ in the opposite
chirality case can be up twice depend on the choose of the parameters $m_c /m_b$ and $c_{12}$.
It is, however, not change our results too much in the later discussion.

\begin{table}[hbt]
\begin{tabular}{|l|ll|l|}
Branching ratio && data &Average \\ \hline \hline
$Br(B\to K^+\pi^-)$&CLEO&$17.2^{+2.5}_{-2.4}\pm1.2$ &$17.3\pm
1.5$\\  &Belle &$19.3 ^{+3.4+1.5}_{-3.2-0.6}$ & \\
 &BaBar&$16.7\pm{1.6}\pm{1.3}$ & \\
$BR(B\to K^-\pi^0)$&CLEO&$11.6^{+3.0+1.4}_{-2.7-1.3}$&$12.1
\pm 1.7$\\&Belle &$16.3^{+3.5+1.6}_{-3.3-1.8}$& \\
&BaBar&$10.8^{+2.1}_{-1.9}\pm{1.0}$ &\\
$Br(B\to \bar K^0 \pi^-)$&CLEO&$18.2^{+4.6}_{-4.0}\pm
1.6$&$17.3 \pm 2.7$\\
 &Belle&$13.7^{+5.7+1.9}_{-4.8-1.8}$& \\
&BaBar&$18.2^{+3.3}_{-3.0}\pm{2.0}$ &\\ $Br(B\to
K^0\pi^0)$&CLEO&$14.6^{+5.9+2.4}_{-5.1-3.3}$&$10.4 \pm 2.7$\\
&Belle&$16.0^{+7.2+2.5}_{-5.9-2.7}$ & \\ &BaBar &$8.2^{+3.1}_{-2.7}\pm{1.2}$&
\end{tabular}
\smallskip
\caption{Experimental
data of $B\to K\pi$ decays from CLEO, Belle and BaBar\protect \cite{12}}
\label{expkpi}
\end{table}

\subsection{ Constraints from $B\to K\pi$}

It is clear from discussions in the previous section $B\to X_s \gamma$ can
constrain $c_{11}^{susy}$,
but still allow large deviations from the SM prediction. Within the allowed regions for
$r_g$, rare $B\to K\pi$ decays can be dramatically
different from the SM predictions. Therefore using
experimental data on $B\to K\pi$ given in Table I \cite{12},
one can further constrain
the allowed regions for $r_g$.

The branching ratio for $B \to K \pi$ with SUSY contribution to
$c_{11}$ can be easily obtained by using $c_{11}=
c^{SM}_{11}(1+r_g e^{i \delta})$ in Eq.(8) for the SUSY
contributions with the same chirality as the SM one, and by using
$c_{11} = c^{SM}_{11}(1-r_g e^{i\delta})$ with opposite chirality case.
To finally obtain the branching ratios, we need to know the form
factors, $F_0^{B\to P}$, the quark masses, the parameters of distribution amplitudes, the CKM parameters
$V_{ub}$, $V_{cb}$ and the phase $\gamma$. There are several
theoretical calculations for the form factors, we will use
$F_0^{B\to \pi} =0.28$ and $R_{\pi K}=0.9$ given in
Ref.\cite{8}. The quark masses $m_b(m_b) = 4.2GeV$, $m_c(m_b) = 1.3GeV$,
$m_s(2GeV) = 110MeV$ and $(m_u+m_d)(2GeV) = 9.1MeV$ will be used for
illustration. For the magnitudes of the CKM parameters we will
use $|V_{cb}|=0.0402$ and $|V_{ub}/V_{cb}| = 0.090$ and treat
$\gamma$ as a free parameter.

To see how the SUSY gluonic dipole interactions can change the SM
predictions in detail, we first study some special cases for
$O_{11}$ with the same chirality as the SM one, and then
consider the combined constraints. For the special cases we
take $r_\gamma$ determined using the central value for $B\to
X_s \gamma$, and take the corresponding $r_g$ with $x_{gq} =
10$ to minimize the effects. For the CP violating phase we
consider three scenarios: a) $r_g e^{i\delta}$ is real and the phase $\gamma$ is the only CP violating
phase; b) the phase $\gamma$ is set to zero and the new
contribution has a phase $\delta$ which can vary from 0 to
$2\pi$; And c) the phase $\gamma$ is fixed at the current best
fit value $\gamma = 66^\circ$\cite{14} in the SM and let the
phase $\delta$ to vary from 0 to $2\pi$.

\begin{figure}[htb]
\begin{tabular}{cc}
\small{(i)$BR(B\to K\pi)$ vs. $\gamma$, SM prediction.} &
\small{(ii)$BR(B\to K\pi)$ vs. $\gamma$, $r_g$ is real.}     \\
\DESepsf(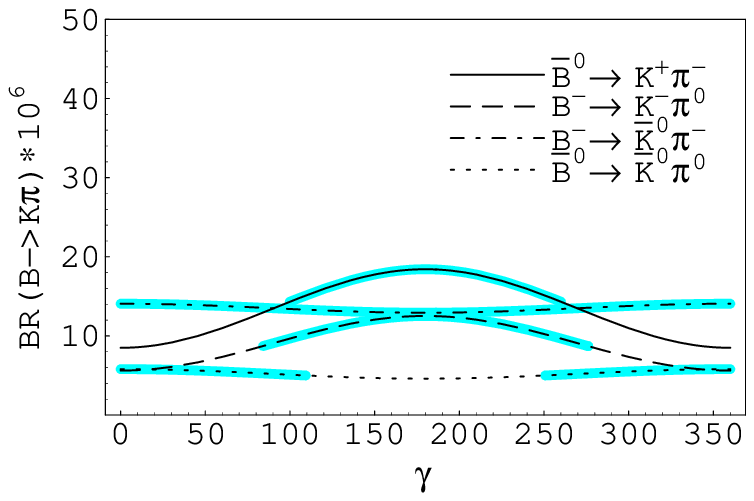 width 8cm) &\DESepsf(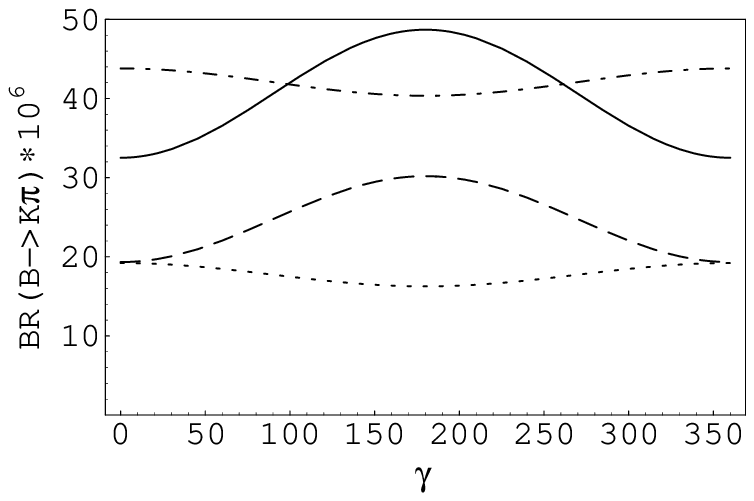 width 8cm)  \\[7mm]
\small{(iii) $BR(B\to K\pi)$ vs. $\delta$, $\gamma=0^\circ$.} &
\small{(iv) $BR(B\to K\pi)$ vs. $\delta$, $\gamma=66^\circ$.} \\
\DESepsf(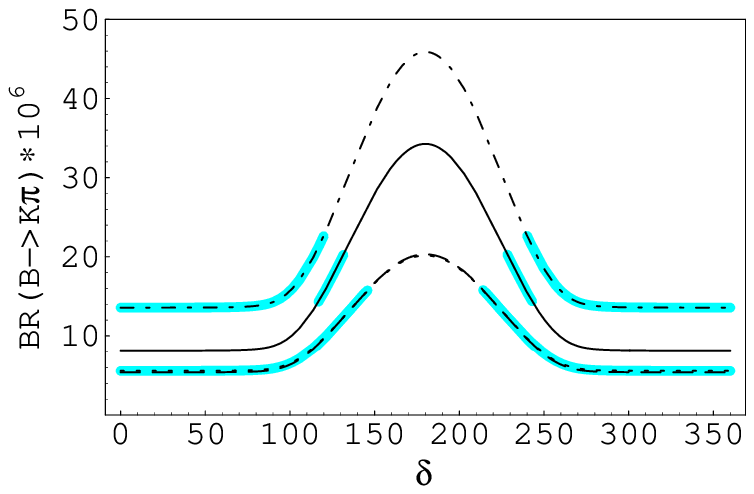 width 8cm) &\DESepsf(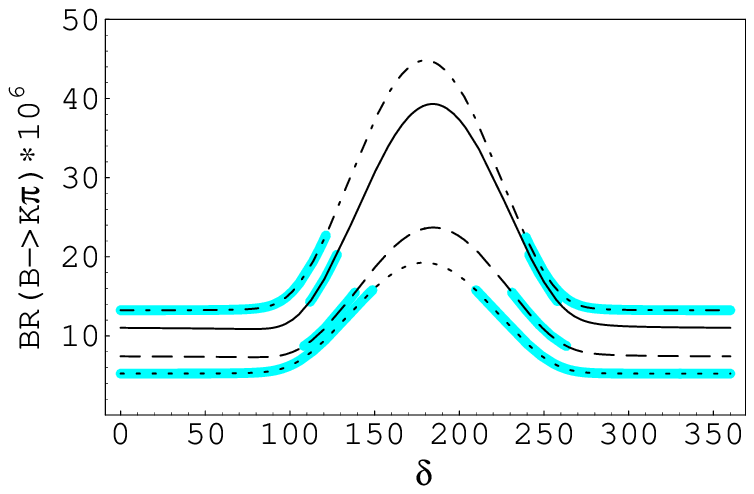 width 8cm)  \\[7mm]
\end{tabular}
\caption {
Branching ratios of $B\to K\pi$. The line segments
with shade indicate the branching
ratios which are within the experimental $2\sigma$ regions.}
\label{special}
\end{figure}

The results are shown in Fig.(\ref{special}).
For comparison, we first show the SM predictions for the four $B\to K \pi$ decays in
Fig.(\ref{special}.i). The $\gamma$ phase seem to be allowed only in the small regions around $100^\circ \sim 109^\circ$ and $251^\circ\sim 260^\circ$,
which are different from the globe fitting\cite{14}.
Of course this situation can be different when different parameters are used\cite{8}.
we will take the value obtained here for illustration.

The situation can be dramatically changed if SUSY gluonic
dipole interactions are included.
In case a), there are two regions of solutions of $r_g e^{i\delta}$, one is
small negative and the other is negative with large magnitude.
The small negative solution $r_g e^{i\delta}$ is too small to influence $B\to
K\pi$ decay. For the large negative $r_g e^{i\delta}$ we
find that the gluino contributions is 5.6
times larger then the SM one, but have opposite sign to the SM one.
In this case, all $B\to K\pi$ decays are too large as can be seen from Fig.(\ref{special}.ii).
If a smaller $x_{gq}$ is used, the
situation is worse. SUSY contribution with no phase ($\delta = 0$)
is not favored.

In case b), when the phase $\delta$
is small, the branching ratios are almost the same as
those in the SM which increase with the phase
$\delta$ to maximal at $\delta = 180^\circ$.
There is a large gap between $B \to K^0 \pi^-$ and $B \to K^+ \pi^-$
shown in Fig.(\ref{special}.iii).
$B \to \bar K^0 \pi^0$ and $B \to K^- \pi^0$ are overlap.
The regions with branching ratios for
$B\to K\pi$ to be within experimental
$2\sigma$ ranges are located between
$117^\circ\sim 120^\circ$ and $240^\circ\sim 243^\circ$.

In case c), the four branching ratios are
shown in Fig.(\ref{special}.iv).
In this case it is, again,
possible to make all decays into experimental $2\sigma$ ranges. The
overlap regions for $\delta$ are located between
$109^\circ\sim 121^\circ$ and $243^\circ\sim 261^\circ$.

From the above discussions, we see that SUSY gluonic dipole interaction can
improve agreement of theoretical predictions and experimental data. Both
CP violating phases $\gamma$ and $\delta$ can affect the branching ratios
significantly.

\begin{figure}
\begin{tabular}{cc}
\small{(a) $\bar B^0 \to K^+\pi^-$} &
\small{(b) $B^-\to K^-\pi^0$ }     \\
\DESepsf(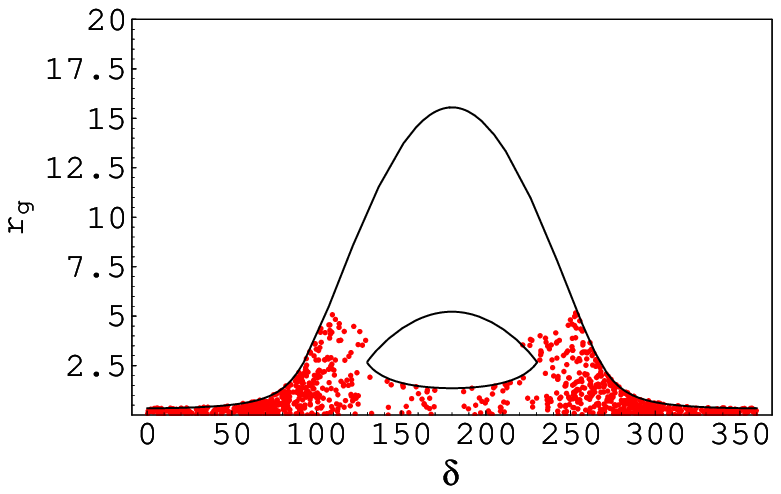 width 8cm) & \DESepsf(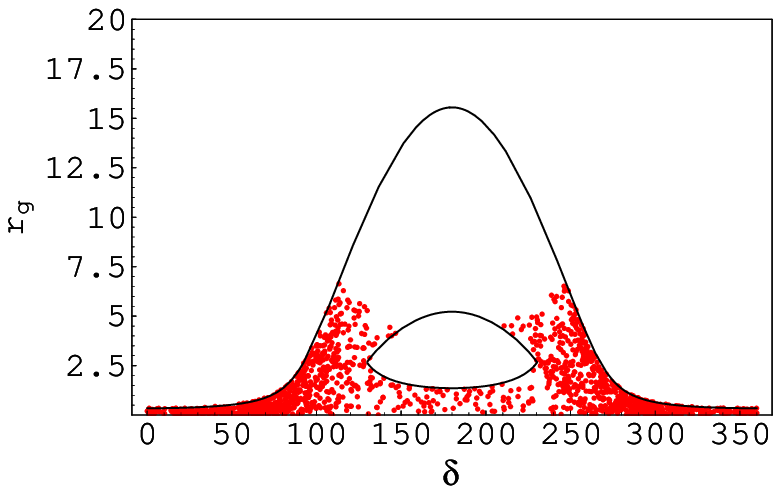 width 8cm)  \\[7mm]
\small{(c) $B^- \to \bar K^0\pi^-$} &
\small{(d) $\bar B^0\to \bar K^0\pi^0$ }     \\
 \DESepsf(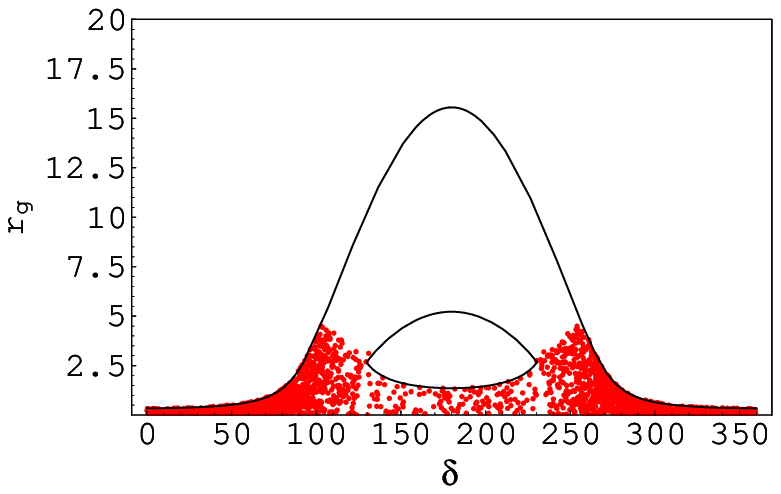 width 8cm) &\DESepsf(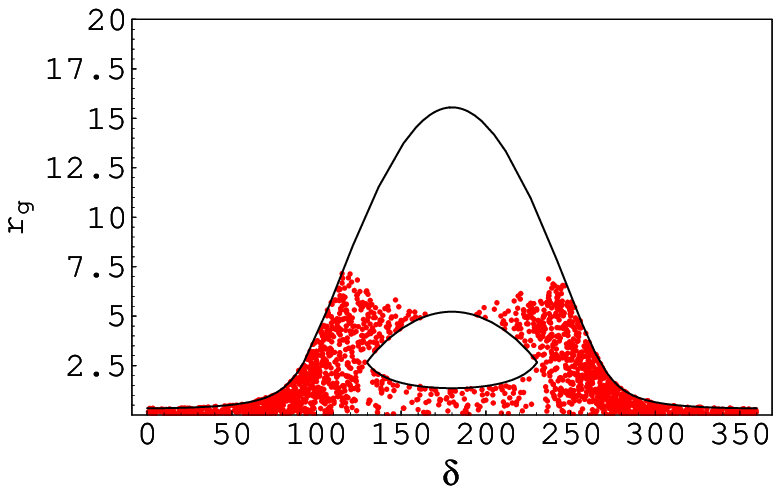 width 8cm) \\[7mm]
\small{(e) All $B \to K\pi$ mode, $0^\circ<\gamma<360^\circ$} &
\small{(f) All $B \to K\pi$ mode, $42^\circ<\gamma<87^\circ$}   \\
\DESepsf(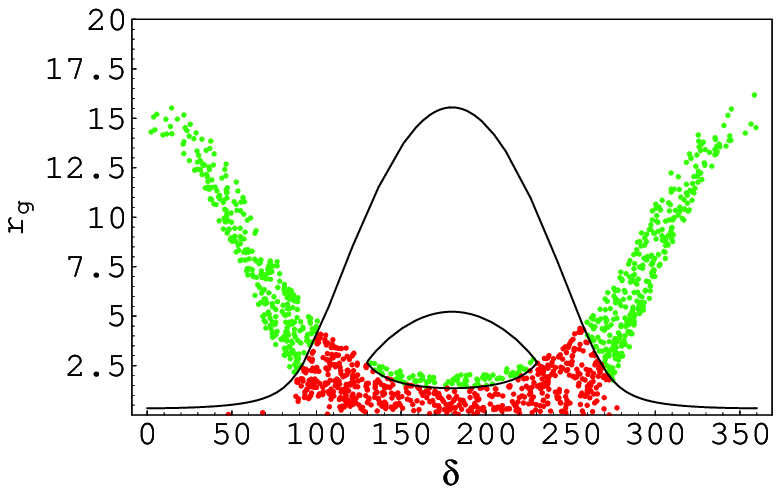 width 8cm)& \DESepsf(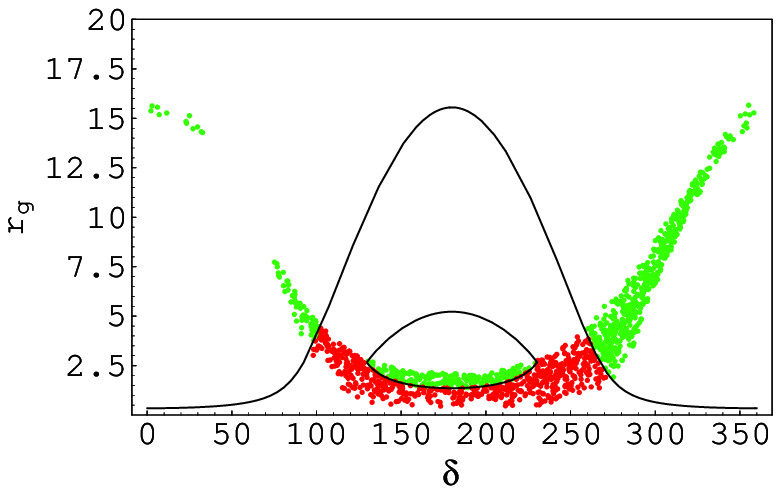 width 8cm) \\[7mm]
\end{tabular}
\caption {
Constraints on $r_g$ using $B\to X_s\gamma$ and $B\to K\pi$ decays.
The solid lines are the boundaries of the constraint from $B\to X_s \gamma$
and the dotted regions are the allowed regions
by data from $B\to K \pi$ at $2\sigma$ level.} \label{kpiconst}
\end{figure}

Since the gluonic dipole interaction can have significant
influence on $B \to K \pi$ branching
ratios, $B \to K \pi$ decays can, therefore, also
be used to constrain SUSY gluonic dipole interaction. To
obtain the restricted regions,
we scan the allowed region of $r_g$ obtained
earlier from $B\to X_s\gamma$ and the
SUSY parameter space $0.1<x_{gq}<10$, $200 \mbox{ GeV}<m_{\tilde g}<1000$ GeV.
The allowed regions for $Br(B\to K \pi)$ within
$2\sigma$ of experimental values
are shown in Fig.(\ref{kpiconst}). In obtaining the allowed regions in
Fig.(\ref{kpiconst}), we treated $\gamma$ as a free parameter varying
in the range $0^\circ\sim 360^\circ$.
Figs.(\ref{kpiconst}.a-d) show the allowed regions from
each $B\to K\pi$ decay.
All four decays constrain $r_g$ to be less than $7.2$.
$\bar B^0\to \bar K^0 \pi^-$ provides
the most powerful constraint on allowed region for $r_g$.
The regions with $r_g>4.4$ are ruled out at $95\%$ C.L..

In Fig.(\ref{kpiconst}.e) we use all four
$B\to K \pi$ decays to
constrain $r_g$ with $\gamma$ varying from $0^\circ$ to $360^\circ$.
$r_g$
is constrained to be less than $4.4$.
$\delta$ between
$0^\circ\sim 80^\circ$ and $280^\circ\sim 360^\circ$ are allowed with small $r_g$ disfavored due to
the reason that it make the branching ratio of $B\to K\pi$ smaller. We also show the constraint with
$42^\circ < \gamma< 87^\circ$, which is the
$95\%$ C.L. of $\gamma$ from the fit of
unitarity triangle in the SM\cite{14} in Fig.(\ref{kpiconst}.f).
In this case we see that the allowed regions of $r_g$ and $\delta$ are
further reduced, $0.3<r_g<4.3$ and
$\delta$ is between $100^\circ \sim 274^\circ$.
Non-zero $r_g$ and $\delta$ fit data better than SM.

\begin{figure}
\centerline{ \DESepsf(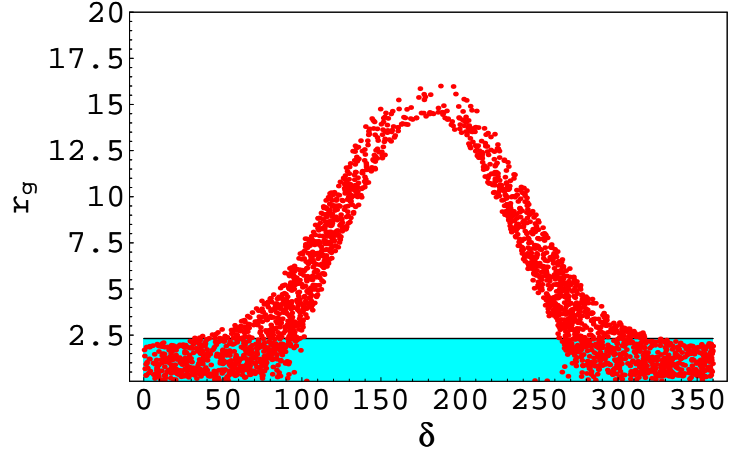 width 10cm)}
\caption {
The allowed regions for SUSY gluonic dipole interaction with
opposite chirality as the SM one.
The dotted regions are the allowed regions from $B\to K\pi$ constraints,
and the region below the dashed line is the upper bound
from $B\to X_s\gamma$ constraint.}\label{LR}
\end{figure}

The constraints on the SUSY gluonic dipole
interactions with opposite chirality to the SM one
can be easily obtained
by using $c_{11} = c_{11}^{SM}(1-r_g e^{i\delta})$ in Eq.(8).
The combined allowed regions on $r_g$,
under the same conditions as for the previous discussions,
are shown in Fig.(\ref{LR}). Data from $B\to K\pi$, again,
can further constrain
the allowed parameter space compared with constraint from
$B\to X_s \gamma$ alone. The region between $100^\circ <\delta<260^\circ$ is not favored.

\section{CP violation in $B\to K\pi$ with susy contributions}

From previous discussions, it is clear that SUSY gluonic dipole interactions
can affect the branching ratios of $B\to K \pi$ significantly.
It is interesting to note that large new CP violating
phase $\delta$ is still allowed which may affect CP violation in these decays.
In this section we study CP violating rate asymmetry

\begin{eqnarray}
A_{asy} = {\bar \Gamma - \Gamma \over \bar \Gamma + \Gamma},
\end{eqnarray}
with SUSY gluonic dipole interactions.

\begin{figure}
\begin{tabular}{cc}
\small{(a) Asymmetry of $\bar B^0 \to K^+\pi^-$} &
\small{(b) Asymmetry of $B^-\to K^-\pi^0$ }     \\
\DESepsf(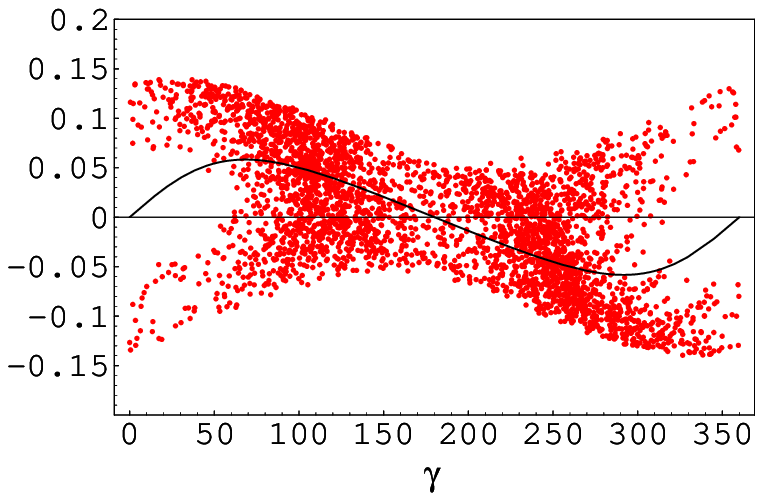 width 8cm) & \DESepsf(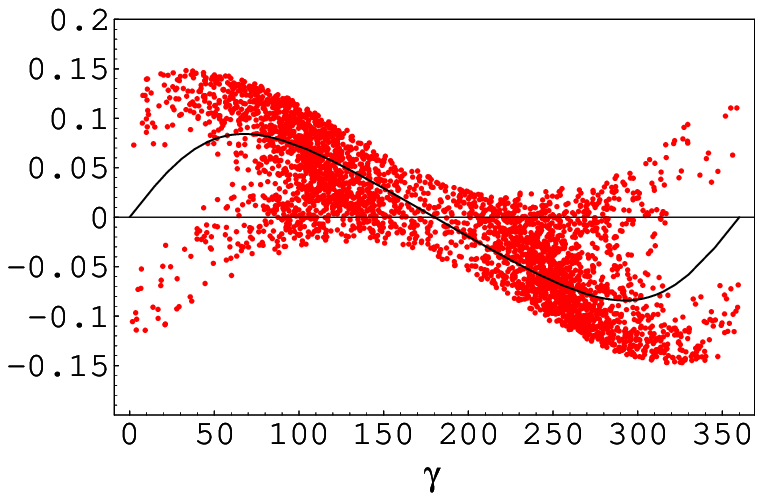 width 8cm)  \\[7mm]
\small{(c) Asymmetry of $B^- \to \bar K^0\pi^-$} &
\small{(d) Asymmetry of $\bar B^0\to \bar K^0\pi^0$ }     \\
 \DESepsf(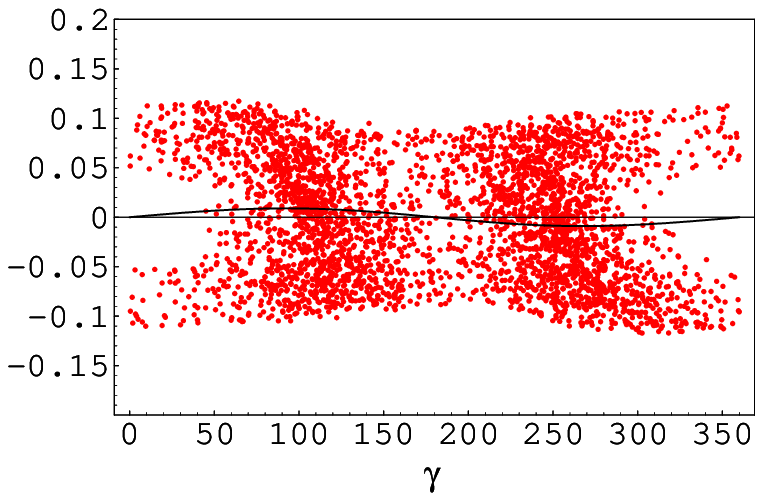 width 8cm) &\DESepsf(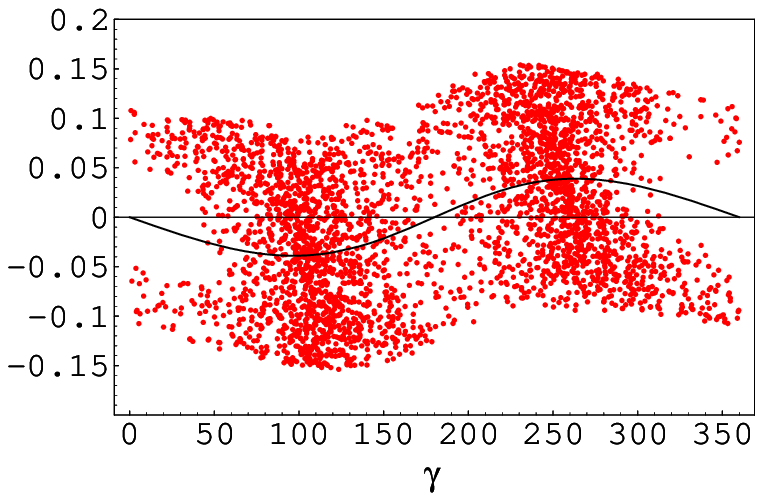 width 8cm) \\[7mm]
\end{tabular}
\caption {
The CP asymmetry $A_{asy}$ for
$B\to K\pi$ decays with SUSY gluonic dipole interaction having the same chirality as the SM one.
The solid curves are SM predictions and
the dotted regions
are predictions with SUSY gluonic dipole interaction.} \label{asym}
\end{figure}

\begin{figure}
\begin{tabular}{cc}
\small{(a) Asymmetry of $\bar B^0 \to K^+\pi^-$} &
\small{(b) Asymmetry of $B^-\to K^-\pi^0$ }     \\
\DESepsf(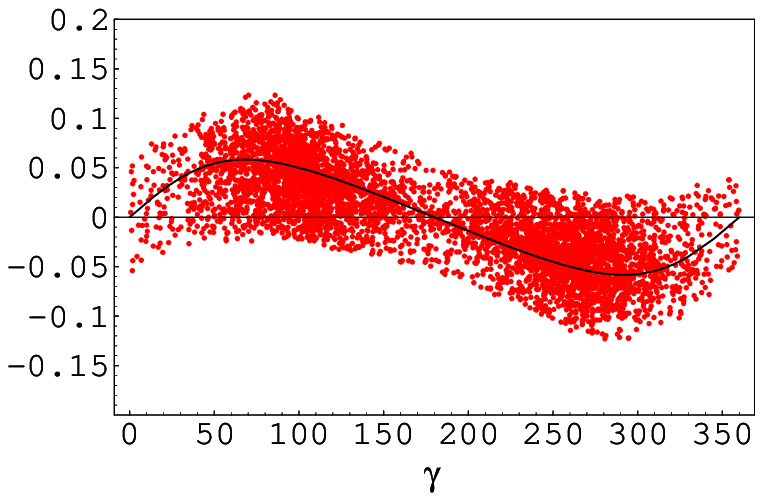 width 8cm) & \DESepsf(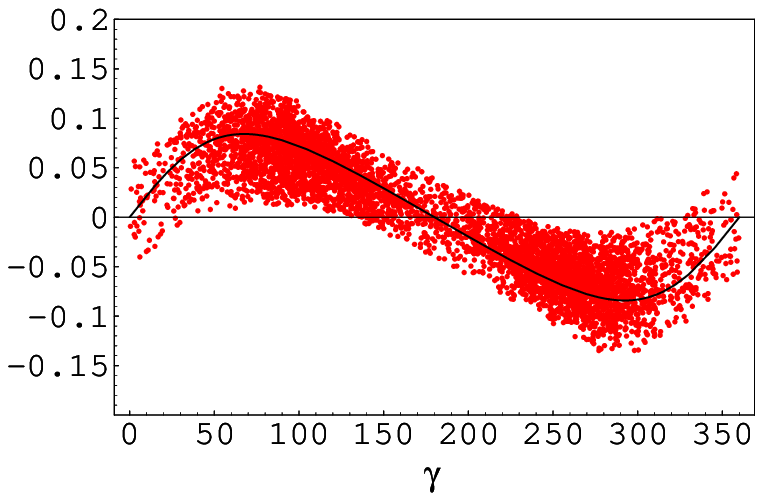 width 8cm)  \\[7mm]
\small{(c) Asymmetry of $B^- \to \bar K^0\pi^-$} &
\small{(d) Asymmetry of $\bar B^0\to \bar K^0\pi^0$ }     \\
 \DESepsf(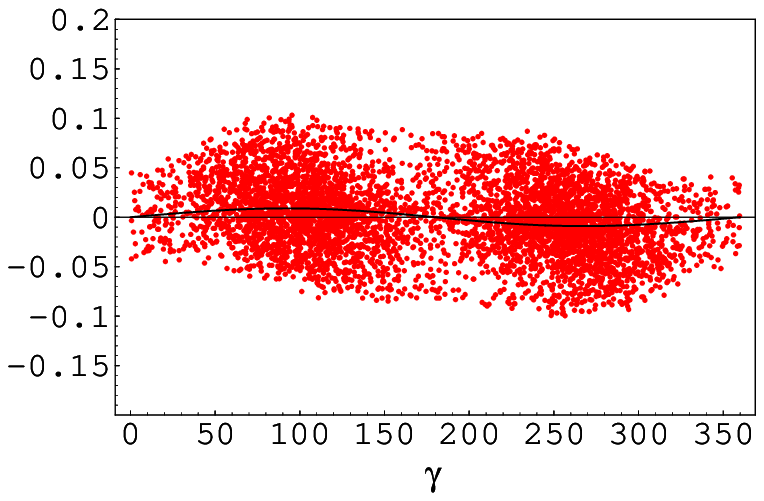 width 8cm) &\DESepsf(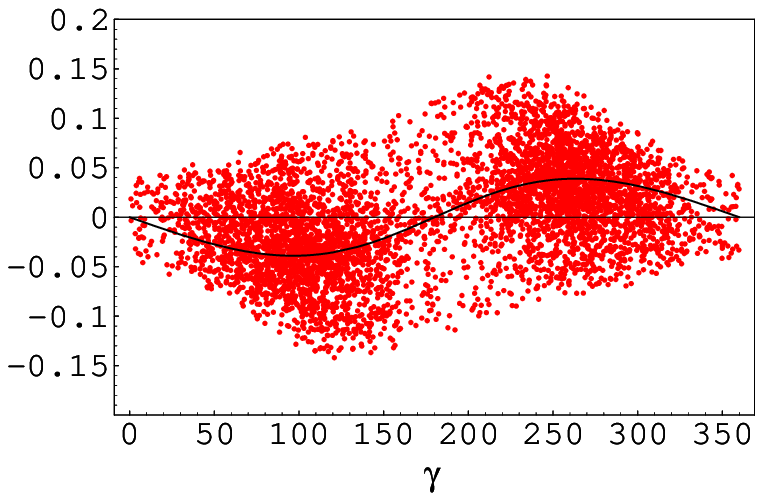 width 8cm) \\[7mm]
\end{tabular}
\caption {
The CP asymmetry $A_{asy}$ for
$B\to K\pi$ decays with SUSY gluonic dipole interaction having the opposite chirality as the SM one.
The solid curves are SM predictions and
the dotted  regions are predictions with SUSY gluonic dipole interactions.} \label{asym1}
\end{figure}

The allowed CP asymmetry $A_{asy}$ for each of the $B\to K \pi$ decay is
obtained by using the allowed regions of parameter space constrained from the
previous section.
The results for SUSY gluonic dipole interactions
with the same chirality as the SM one and opposite
one are shown in Fig. (\ref{asym}) and Fig. (\ref{asym1}), respectively.
The reference
SM predictions as a function of $\gamma$
are shown as solid lines in Fig.(\ref{asym}).
When SUSY gluonic dipole contributions are included, the asymmetries can be dramatically
different because $r_g$ and $\delta$ both can be large. The scattered dots in
Fig. (\ref{asym}) above
and bellow the SM predictions correspond to
the regions on $\delta>180^\circ$ and $\delta<180^\circ$ in Fig.(\ref{kpiconst}e), respectively.
We see clearly that the predictions can be much larger than the SM
predictions.
For example,
with $r_g =3$, $\delta =120^\circ$ and $\gamma = 66^\circ$,
the branching ratios
for $K^+\pi^-$, $K^-\pi^0$, $\bar K^0\pi^-$, and $\bar K^0\pi^0$,
are predicted to be $(16.8,\;10.7,\;21.8,\;9.2)\times 10^{-6}$
which are within the $2\sigma$ region of data,
and the asymmetries
are predicted to be $-0.04$, $0.00$, $-0.06$, $-0.10$, respectively.
With $r_g =3$, $\delta =250^\circ$ and $\gamma = 66^\circ$, the branching ratios are
$(19.0,\;12.1,\;20.0,\;8.1)\times 10^{-6}$,
and the asymmetries are $0.11$, $0.12$, $0.09$, $0.07$, respectively.

For SUSY gluonic dipole interaction
with the opposite
chirality, the allowed CP asymmetry $A_{asy}$ for each of the
$B\to K \pi$ decay
are similar as the above examples.
The results are shown in Fig.(\ref{asym1}).

A particular interesting case is that for $B^- \to \bar K^0 \pi^-$.
The SM predicts that
$|A_{sym}(B^- \to \bar K^0 \pi^-)|<1\%$. In the \cite{8}, considering all kinds of uncertainty, $|A_{sym}(B^- \to \bar K^0 \pi^-)|$ is still not larger than $2\%$.
With SUSY dipole interaction, $|A_{asy}(B^- \to \bar K^0 \pi^-)|$
can be as large as $10\%$. Observation of CP violation in $B^-\to \bar K^0\pi^-$ significantly
large than SM prediction is an indication of new physics.

\section{conclusions}

In this paper we have studied the contributions of
SUSY gluonic dipole interaction to $B\to
K\pi$ decays.
We found that SUSY gluonic dipole interactions can affect $B\to K \pi$ significantly.
QCD improved factorization calculation is lower than the current experimental
branching ratio for $B^0\to K^0 \pi^0$.
If experimental data will be further
confirmed, this is an indication
of new physics. We found that SUSY dipole interactions can
improve the situation.  All
four measured $B\to K \pi$ decays can be in
agreement with theoretical calculations with SUSY gluonic dipole interactions
and at the
same time satisfy constraint from $B \to X_s\gamma$.

Present data from $B\to K \pi$ can also further constrain the allowed parameter space for SUSY gluonic dipole interactions.
A large portion of the parameter space allowed by $B\to X_s \gamma$
are excluded by $B\to K \pi$ data.
SUSY gluonic dipole interaction coefficients $r_g$ and $\delta$
are constrained to be into two narrow regions for
both types of dipole chiralities.

Constraints from $B\to X_s \gamma$ and $B\to K \pi$ still allow
large new CP violating
phases in the SUSY gluonic dipole interactions.
This allows very different predictions for CP
asymmetries for $B\to K\pi$ decays.
In particular, with SUSY gluonic dipole interactions the value of
CP asymmetry for
$B^-\to \bar K^0 \pi^-$ can be as large as $10\%$
which is much larger than the
SM prediction of less than 1\%. This can provide an important test of
new physics beyond the SM. CP violation in other modes can also be larger
than the SM predictions.

Finally we would like to comment that with SUSY gluonic dipole contributions,
some relations predicted in the SM may be violated. One such example is the
rate difference, defined as $\Delta = \bar \Gamma -\Gamma$. In the Standard
Model, one has $\Delta (\bar B^0 \to K^- \pi^+) =-(f_k/f_\pi)^2\Delta(\bar B^0
\to \pi^+\pi^-)$\cite{15},if kaon and $\pi$ have the same wave function
distribution amplitude.With different distribution amplitudes the relation is violated at
$10\%$ level.  In obtaining this an important property of the
CKM matrix element $Im(V_{ub}V_{us}^*V_{tb}^*V_{ts}) = -Im(V_{ud}V_{ud}^*
V_{tb}^*V_{td})$ has been used. Since the SUSY contributions are proportional to
$\delta_{LR,RL}^{bs}$ for $B\to K \pi$ decays, and proportional to
$\delta_{LR,RL}^{bd}$ for $B\to \pi\pi$ decays. In general
$\delta^{bs}_{LR,RL}$ and $\delta^{bd}_{LR,RL}$ are not related,
the contributions
from SUSY will break the relation $\Delta(\bar B^0\to K^- \pi^+)
= -(f_K/f_\pi)^2 \Delta(\bar B^0 \to \pi^+\pi^-)$. We have checked with numerical
calculations and found that indeed
when SUSY contributions are included
there are regions of parameter space where the relation mentioned is
badly violated. Experimental measurements
of these rate differences can also serve
as tests of new physics beyond the Standard Model.

\acknowledgements
We thank M. Bebeke, H.-Y. Cheng and A. Kagan for useful suggestions.
This work was supported in part by NSC
under grant number NSC 89-2112-M-002-058, by the NCTS and by the
MoE Academic Excellence Project 89-N-FA01-1-4-3.

\end{document}